\documentstyle[preprint,12pt,aps,prb]{revtex}
\tightenlines
\begin{document}
\newcommand{\nvo}{NaV$_{2}$O$_{5}$}
\newcommand{\cvo}{CaV$_{2}$O$_{5}$}
\newcommand{\cm}{cm$^{-1}$}
\newcommand{\cgo}{CuGeO$_3$}
\newcommand{\mvo}{MgV$_{2}$O$_{5}$}
\title{Charge-ordering phase transition and order-disorder
effects in the Raman spectra of NaV$_2$O$_5$}
\author{M.J.Konstantinovi\'c $^{\ast}$}
\address{Simon Fraser University, Physics Department, 8888 University 
drive, Burnaby, B.C. V5A1S6 Canada}
\address{Max-Planck-Institut f{\"u}r Festk{\"o}rperforschung, Heisenbergstr. 1, 
D-70569 Stuttgart, Germany} 
\author{M.Isobe and Y.Ueda}
\address{Institute for Solid State Physics, The University of Tokio, 
7-22-1 Roppongi, Minato-ku, Tokio 106, Japan} 

\maketitle

\begin{abstract} 

In the ac polarized Raman spectra of NaV$_2$O$_5$ we have found anomalous
phonon broadening, and an energy shift of the low-frequency mode as a
function of the temperature. These effects
are related to the breaking of translational symmetry,
caused by electrical disorder
that originates from the fluctuating nature of the
V$^{4.5+}$ valence state of vanadium.
The structural correlation length, obtained from comparisons between
the measured and calculated Raman scattering spectra, diverges at $T<$ 5 K,
indicating the existence of the long-range charge order at very low
temperatures, probably at $T=0$ K.
\end{abstract}


The investigation of the physical phenomena associated to
the low-dimensional magnetic
structures, like a spin-Peierls (SP) transition discovered in the inorganic 
compound CuGeO$_3$ \cite {a2}, are of great importance for better 
understanding of strong electron correlations. A very interesting 
interplay between spin and charge dynamics results in a phase transition 
discovered in NaV$_2$O$_5$ \cite {a1a,a2a}.
At higher temperatures ($T>34 K$) the magnetic susceptibility of {\nvo} is in
excellent agreement with the Bonner-Fisher curve for the one-dimensional
Heisenberg antiferromagnet. At low temperatures, however, the susceptibility
decreases rapidly to zero \cite {a1a} suggesting a SP transition at T=34 K. On 
the 
other hand, the temperature dependent nuclear magnetic resonance (NMR) spectra, 
\cite {a3} showed a change of the vanadium valence across the phase 
transition, from uniform V$^{4.5+}$ to two different V$^{4+}$ and V$^{5+}$ 
states. These measurements \cite{a1a,a3} gave direct evidence for the 
charge 
ordering (CO) phase-transition scenario in {\nvo}. A structural analysis, \cite 
{a4} and 
the polarized Raman and infrared (IR) spectra of NaV$_2$O$_5$ \cite {a5}, are 
also consistent with the existence of uniform vanadium valence in the 
high-temperature 
phase.
However, despite extensive experimental and theoretical effort,
no consistent picture has yet emerged for the low-temperature
phase
\cite {a6,a8,a81,a82,a83,a84,a85,a86,a87,a88,a89,a90,a91,a92}.
None of the models
proposed to date, have explained fully the two intimately connected issues: the 
nature of the phase transition and the low-temperature ground state. Raman 
spectroscopy is a powerful method for the study of dynamics of solids and it can 
be used 
to address these issues. Again, despite a considerable amount of data, the 
Raman spectra of \nvo\ are still not completely understood \cite {a5,a12}. This 
holds for the vibrational modes as well as 
for possible magnetic excitations. 

In this report, we present the study of the vibrational modes in the Raman 
spectra of \nvo\, and analysis of the effects associated with non conservation 
of the
light-scattering wave vector selection rule {\bf k}=0. We argue that anomalous 
phonon
broadening and frequency shifts, that are observed in the ac spectra as a 
function of the 
temperature, are caused by strong electrical disorder due to the fluctuating 
nature of the V$^{4.5+}$ valence state of vanadium. The temperature dependence 
of 
the structural correlation length is obtained from the comparison between 
measured and calculated Raman scattering spectra, and used to characterize the 
phase transition in NaV$_2$O$_5$. This analysis shows, that the phase transition 
at $T_c=34 K$ corresponds to the onset of the short-range electron correlations, 
with a true long-range order (CO) at $T \sim $0. Thus, the charge order is not 
static below Tc, in contrast to the conclusions of Lohmann et al \cite {a92}.

Polarized Raman scattering experiments were performed on \nvo\ single crystals 
(size $\sim$\ $1\times3\times1$~mm$^3$ along {\bf a}, {\bf b}, and {\bf c}
respectively)
prepared as described in Refs. 2 and 3. As an excitation source we used the 
514.5 nm 
laser line from an Ar$^+$ ion laser.  The beam, with an average power of 5~mW, 
was 
focused (spot diameter $\sim80\mu$m) on the (001) and (101) surfaces of the 
crystals. The spectra were measured in a backscattering geometry using a {\sc 
dilor} triple monochromator equipped with a LN$_2$ cooled CCD camera. 
NaV$_2$O$_5$ crystallizes above T$_c$ in the orthorombic centrosymmetric space 
group Pmmn (D$_{2h}^{13}$), with two molecules in the unit cell of a size:  
a=1.1318 nm, b=0.3611nm and c=0.4797 nm.  Each vanadium atom, in the 
average valence state 4.5+, is surrounded by five oxygen atoms forming 
VO$_5$ pyramids.  These pyramids are mutually connected via common edges and 
corners to form layers in the ab plane, see Fig. 1. The Na atoms are situated 
between 
these layers as intercalants. The structure of NaV$_2$O$_5$ can also 
be described as an array of parallel ladders (running along the {\bf b} 
direction) 
coupled in a trellis lattice (each rung is made of a V-O-V bond).

The vibrational properties of \nvo\ are studied in great details by measuring 
the Raman spectra only from (001) surface \cite {a5,a12}. In our previous paper
\cite {a82} we observed an interesting low-frequency structure in the
ac scattering
geometry which, as we will show below, exhibits an unusual temperature 
dependence.
The polarized Raman scattering spectra from (001) and (101) planes, at 
temperatures above and below phase-transition temperature, are presented in 
Fig. 2. The {\bf b} axis of the crystals is set to be parallel to the laboratory 
H horizontal) axis. Thus the VV (V equals vertical) polarized scattering geometry from 
the (101) plane gives the mixture of aa, cc, and ac contributions (upper spectra 
in Fig. 2). The VV polarized configuration from the (001) plane gives only aa 
contribution (lower spectra in Fig. 2). The corresponding HH spectra (gives
the bb contribution for both planes) are found to be identical. In this way we 
were 
able to determine the ac contribution at very low frequencies, otherwise 
impossible to get. For example, the low quality of the (010) surface, prevent us 
from direct measurements of the low frequency scattering in the ac geometry.

In addition to the phonon modes, and a continuum centered at 650 $cm^{-1}$ which 
belongs to the aa channel, we found three modes in the VV spectra 
obtained from the (101) planes (Fig. 2), that represent the ac contribution. 
Their 
detailed temperature dependencies are shown in inset of Fig. 2. The lowest 
frequency ac mode is centered at about 107 $cm^{-1}$. The mode is asymmetric 
with a frequency cutoff around 120 {\cm} (Fig.2 and Fig.3b).
As the temperature is decreased below $T_c$, the mode becomes more symmetric,
and hardens by amount of 10 {\cm} [Fig. 2 and Fig. 3b]. The integrated intensity 
of
this
feature increases rapidly by the increasing temperature below $T_c$,
and becomes constant for $T > T_c$ (Fig. 4).
The temperature
behavior of the frequency and the integrated intensity is similar to what is 
usually observed for a two-magnon mode \cite {a13}. However, the two-magnon 
origin of this mode can be ruled out since its energy
(117 {\cm}) below the phase transition
temperatures \cite {a85} is much smaller then $2\Delta_s \sim 160$ {\cm}.
The additional argument for excluding the two-magnon scattering process comes
from observation of the mode at temperatures as high as $10 \times 
T_c$. At these temperatures the two-magnon mode should not be visible in the 
Raman spectra \cite {a13}. A one-magnon scattering process, and other
magnetic-related scattering mechanisms, can be also eliminated since 
we did not find any change in the spectra in the magnetic fields up to 12 T.

Our analysis shows that this structure might be related to the low frequency 
phonon density of states, due to the breakdown of the {\bf k}=0 
conservation rule in the light scattering process. In \nvo\, the V$^{4.5+}$
valence state of vanadium causes the random nature of the coupling between the
atomic displacements and the fluctuations of the dielectric susceptibility.  
Such randomness actually appears because of irregular atomic 
bonding. The visualization of randomness in \nvo\ is schematically presented in 
Fig. 1 and it can be referred as an "electrical disorder", \cite {a14}. Imagine 
that we "freeze" the electrical configuration of the charges in the rungs of the 
ladders, at certain time in the high-temperature phase. We find random electron 
configurations among rungs, Fig.1b. The average value for all rung 
configurations gives the V$^{4.5+}$ in which the electron is shared
by the V atom at the each end of a rung.
Below the phase transition temperature the charges
start to order, (V$^{4.5 \pm \alpha}$, $\alpha \neq 0$) Fig. 1c, reaching 
complete "zig-zag" \cite {a16} order at T=0 ($\alpha = 0.5$), Fig. 1d. The "zig-
zag" phase is presented as a real low-temperature geometry, even though there 
are 
some other proposed CO configurations \cite {a16a,a81}. It will be evident from 
our results that we were not able to discriminate among the various
possible CO patterns,
and Fig.1 must be regarded only as illustrative of the electrical 
disorder.
However, the "zig-zag" charge order is consistent with observed phonons
in the Raman and IR spectra of {\nvo} \cite {a82}.
If so, even for the
perfect plane-wave phonon, for example an acoustic or optical phonon with a 
finite 
wave vector {\bf k}, disorder of the atomic coupling allows inelastic scattering 
of light from this mode \cite {a15}. Then, the light scattering is expected to 
be proportional to the phonon density of states properly weighted by the 
coupling constant which is in fact a function of $\omega$.

The light-scattering cross section is proportional to the
Fourier transform of the correlation function of the polarizability
fluctuations:

\begin{equation}
I_{i,j}(\omega) \sim \int \int dt d({\bf r}_1-{\bf r}_2)
e^{i\omega t - i{\bf k}({\bf r}_1-{\bf r}_2)}
<\delta \chi^{*}_i({\bf r}_1,t) \delta \chi_j({\bf r}_2,0)>,
\end{equation}
where ${\bf k}={\bf k}_I-{\bf k}_S$ and $\omega=\omega_I-\omega_S$ are
wavevector and frequency of phonon(s) that participate in the light scattering 
process. The $<...>$ denotes the equilibrium expectation value. 

For the ideal crystals, both energy and wavevector conservation rules are 
fulfilled, and the first-order Raman intensity of phonons is proportional to two 
delta functions, $\delta(\omega_I-\omega_S-\omega) \cdot \delta ({\bf k}-q)$.

In disordered crystals the {\bf k}=0 selection rule is broken due to lack of 
the translational invariance, and the Stokes part of the Raman scattering 
intensity is 
$I(\omega)\omega / [1+n(\omega)] \sim \sum_{j} C_j(\omega) g_j(\omega)$. 
The $g_j(\omega)$ is a density of states for the band j, and the $n(\omega)$ is 
Bose distribution function \cite {a14}. $C_j(\omega)$ describes the coupling of 
the light and the vibrational mode $\omega$.
In a continuum description \cite {a14} the
fluctuations of the susceptibility result from the elastic strain field 
$e_j({\omega ,\bf r})=ke_j(\omega)exp(i{\bf k}{\bf r})$
of the phonon (plane wave), which is modulated by static fluctuations 
$\delta p_{ik}({\bf r})$ of the elasto-optical constants: 
$\delta \chi_i({\bf r})=-\epsilon^2 /(4\pi)[p_{ij}+\delta p_{ij}({\bf 
r})]e_j({\bf r})$. In this way, the coupling constant 
$C(\omega) \sim k^2 \int dr exp(i{\bf k}{\bf r})F({\bf r})$ 
is expressed as a correlation function $F({\bf r})$ of the fluctuations $\delta 
p$ 
of the elastoptical constants, which characterize the electrical disorder.
The form of the correlations is usually postulated to be either exponential 
damping 
$exp(-r/l_c)$; or Gaussian damping $exp(-r^2/l_c^2)$, where $l_c$ is correlation 
length. In the case of \nvo\, we assume Gaussian damping to describe the 
electrical correlations. The correlation length may be defined as a length over 
which electrons in neighboring rungs "see" each other. In fact, this is just a 
positional (structural) correlation length. One simplified picture, where the 
intersite Coulomb interactions are "switched on" at T=34 K
has been previously suggested \cite {a16}.

Thus, the coupling constant is $C(\omega) \sim k^2 e^{-k^2l_c^2/4}$.
The $\omega$ dependence of $C(\omega)$ comes from the dispersion relation 
between the 
frequency and the wavevector [for example, if $\omega =ck$ then 
$C(\omega)={\omega/c)^2 exp[-(\omega/c)^2l_c^2/4}$].
The same type of the correlation function and the coupling constant have been
obtained by Martin and Brenig \cite {a17}
in their analysis of Brillouin scattering in the
amorphous solids.

Finally, the normalized Raman intensity is:

\begin{equation}
I(\omega) \frac{\omega}{1+n(\omega)} \sim
\sum_{\bf k}k^2e^{-k^2l_c^2/4} \delta [\omega -\omega({\bf k})],
\end{equation}
where $\omega({\bf k})$ is a phonon dispersion.
If we confine our analysis to the energy range of the acoustic
or/and low frequency optical phonons
($\omega \leq 150$ {\cm}) the Raman spectrum
in the high-T phase
is influenced by two contributions: acoustic or optic
phonon density of state contribution,
$\sum_{k} \delta [\omega -\omega({\bf k})]$
and the coupling function $C(k)$.
Since dispersion curves of the phonons have not yet been measured,
we are forced to make assumption about which phonons are involved in
the light scattering process. There are two possibilities:

(a) Broad
feature corresponds to the acoustic phonon with a zone-boundary energy of
117 {\cm}.
Bellow $T_c$ this mode is introduced to {\bf k}=0 by zone folding
effect. The x-ray diffraction experiments showed the existence of 
superlattice reflections below $T_c$ with a lattice modulation
vector {\bf q}=(1/2, 1/2, 1/4) \cite {a85}.
In this case the strong anomaly of the elastic
constants is expected at $T_c$ and indeed observed by
Schwenk {\it et al.} \cite {a84}.

(b) The mode corresponds to
low-$\omega$ optical phonon, also allowed by the symmetry of the
low-T phase, with energy that decreases as a function of the wavevector.
According to the lattice dynamical calculations \cite {a19} a
good candidate for that phonon could be the low-$\omega$ $B_{2g}$ phonon
(active in ac polarized geometry), with  Na vibrations  mainly
along the {\bf a} axis.
By examining the Raman spectra it is difficult to conclude which one of these 
two
assumptions is more appropriate, because of the strong quasi elastic
background at low frequencies. However, this choice does not critically
influence our conclusions, 
and for the sake of simplicity we assume the cos k form
of the phonon dispersion, $\omega=\omega_0 cos k/2$, $\omega_0 = 117$ {\cm},
$k \in [0, \pi /a]$.

The calculated Raman spectra are presented in Fig. 3a and compared
with measured ones, Fig 3b. Both the shift and the broadening
of the mode are in good agreement with the experiment. The spectra are obtained
by varying just one parameter $l_c$, evaluating equation 2
in one dimension, and by taking the values of k and $l_c$ in appropriate units
of a-lattice constant. 
By increasing temperature (from T=0), disorder is introduced and the 
contribution 
of the
$C(k)$ in the spectra becomes important. The increase of the degree of
disorder is produced
by the increase of the $k \neq 0$ contributions, directed with $C(k)$. 
Therefore, the broadening of the mode and its energy shift towards lower
energies are produced by decreasing $l_c$ which is a measure of the
degree of disorder; complete disorder is characterized with $l_c \sim$
a-interatomic distance and long range-order with $l_c= \infty$. 
The long-range order solution, $l_c= \infty$
and $C(k)=0$ of equation 2, gives the vanishing of the Raman
intensity. This is not unphysical.
It is just telling us that one has to analyze the Raman spectra at T=0
using equation 1 instead of equation 2.

The peak position of the mode as a function of the temperature is shown 
in Fig. 4a. The circles (full lines) represent the experiment (theory). 
Since in our calculation of the Raman spectra the temperature does not 
enter as a parameter, the temperature dependence of the peak maximum 
is included only through the temperature dependence of the correlation length. 
It is generally expected that correlation length changes with temperature,
and this dependence can be obtained by examining the phonon frequency change
close to the phase transition-temperature. 
The best agreement between the measured and the calculated spectra 
is obtained by assuming a quadratic relation between the correlation 
length and the temperature, $1/l_c \sim T^2$.
Therefore, the peak position can be used as a measure of
disorder. Accordingly, a similar temperature dependence
is expected for the order parameter $\alpha$.
Furthermore, we present the comparison between the temperature dependence of the
measured and calculated integrated intensities, Fig. 4.b. Here, we also find
very good agreement between theory and experiment using
the same quadratic relation between $1/l_c$ and T, and by introducing
a non zero intensity offset at T=0 K ($I(T=0)/I(T=100)=0.3$).

The temperature dependence of the correlation length is presented in inset 
of Fig. 4. 
The particular value of lc that is used to calculate the Raman spectra, 
does not have deeper physical meaning because of an arbitrary factor 
in the exponent of equation 2. But its temperature dependence does. The 
correlation length has a approximately constant value above $T_c$
and increases below $T_c$. This signals
that the critical temperature $T_c=34$ K represents the onset of short-range
electron correlations. As the temperature is lowered below $T_c$,
the correlation length rapidly increases indicating the existence
of a possible singularity at temperatures below T=5 K. In
this case the divergence of the correlation length corresponds to the appearance
of the true long-range charge order at very low temperatures (probably at
T=0).
The short range electron correlations could correspond to intersite
Coulomb
interaction effects \cite {a16}, which become important below 34 K,
but the electron correlations should also persist in some form at
temperatures above 34 K.
Following the same arguments, the change of the spectral shape above 100 K
(it is also found that some IR spectral changes occurs around 100 K, \cite {a6})
could be a consequence of an additional phase transition,
 magnetic in origin for example. 

In conclusion, we have presented the evidence for the
existence of translational symmetry breaking
effects in the Raman spectra of \nvo.
Non conservation of the light-scattering wave vector selection rule,
${\bf k} \approx 0$,
is caused by strong electrical disorder due to the fluctuating nature of 
the valence of vanadium.
The temperature
dependence of the structural correlation length has been obtained from the
comparison between the measured and the calculated Raman scattering spectra.
This suggests that the phase transition at $T_c=34 K$, represents the onset
of the short range electron correlations, with true
long-range order developing at $T \sim $0. 

{\bf Acknowledgments}

MJK thanks to J.C.Irwin, I.Herbut and P.H.M. van Loosdrecht
for helpful discussions and comments.
This work is supported by Natural Sciences and Engineering Research
Council of Canada. MJK also thanks to MPI-FKF Stuttgart, Germany for partial
financial support.

$^{\ast} mkonstan@sfu.ca$

\begin{figure}
\caption
{a) Schematic representation of the NaV$_2$O$_5$ crystal structure in the
(001) plane. The full circles represent oxigens, the small (large) circles
represent vanadium (sodium) atoms. Schematic representation of the electrical
 disorder (order) at
b) $T>T_c$, c) $T<T_c$, d) T=0 in the vanadium-oxygen
ladders of NaV$_2$O$_5$. Each rung in the ladders represents the
V-O-V bond with one electron (black dot).}
\label{fig1}
\end{figure} 
\begin{figure}
\caption
{The polarized Raman scattering spectra of NaV$_2$O$_5$ at 300 K (thin
line) and at 10 K (thick line),
measured from (001) surface (lower spectra) and (101) surface (upper spectra).
Arrows represent modes associated with the ac polarized contribution, see text.
Inset: The temperature dependent (aa+cc+ac) Raman scattering spectra.}
\label{fig2}
\end{figure} 
\begin{figure}
\caption
{The a) calculated and b) measured Raman scattering spectra in 50 to
130 $cm^{-1}$ frequency range at various temperatures.} 
\label{fig3}
\end{figure} 
\begin{figure}
\caption
{The a) frequency and b) normalized integrated intensity of the
117 $cm^{-1}$ mode
as a function of the temperature. The circles (full lines) represent
the measurements (theory). The dashed line is guide to an eye.
Inset: The temperature dependence of the correlation length. }
\label{fig4}
\end{figure} 

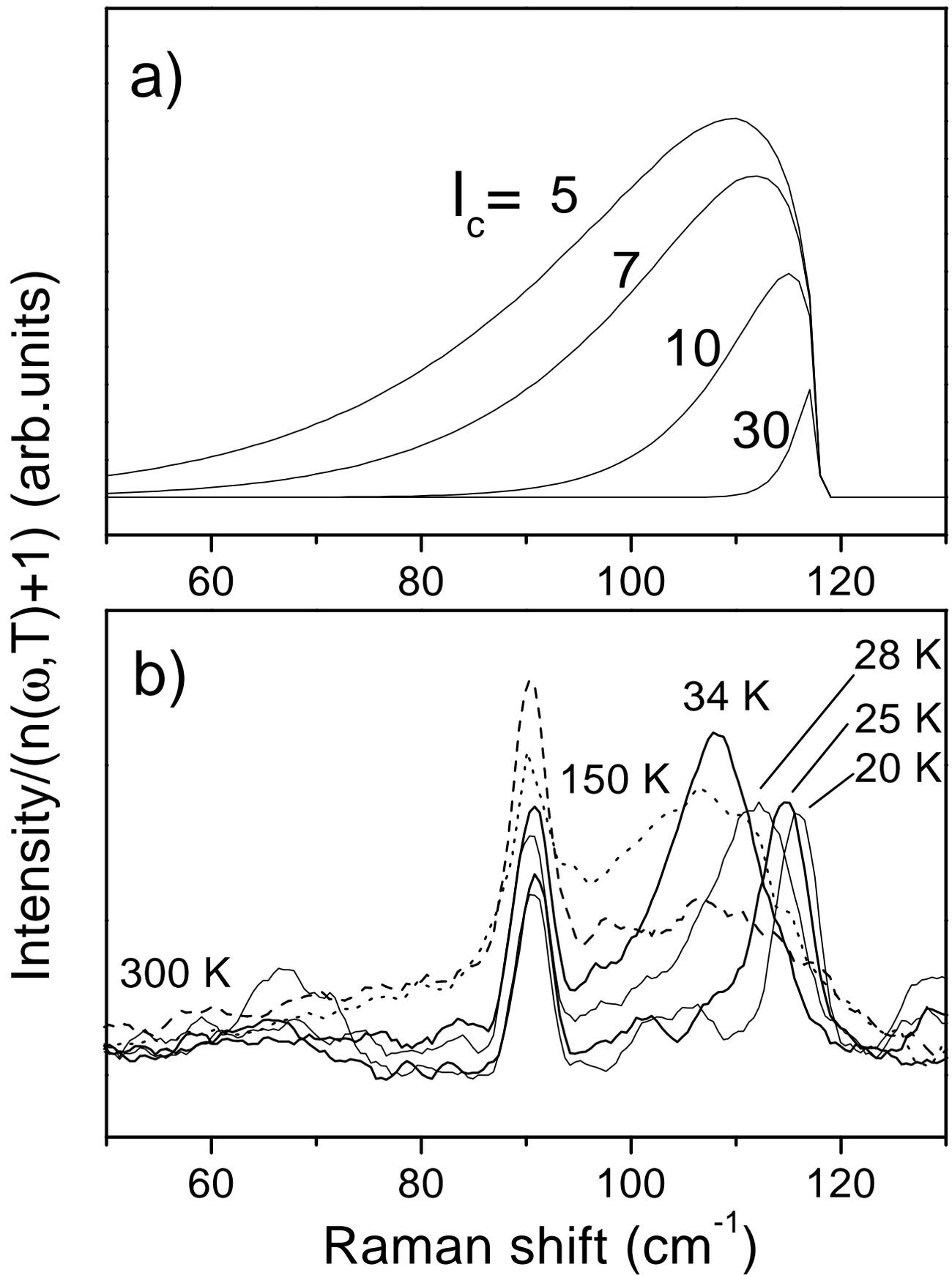
\end{document}